# Impact of Geometric Uncertainty on the Computation of Abdominal Aortic Aneurysm Wall Strain

Saeideh Sekhavat[1[0009-0006-9280-2781]], Mostafa Jamshidian[1*[0000-0002-5166-171X]], Adam Wittek[1[0000-0001-9780-8361]], Karol Miller[1[0000-0002-6577-2082]]

[1] Intelligent Systems for Medicine Laboratory, The University of Western Australia, Perth, Western Australia, Australia
mostafa.jamshidian@uwa.edu.au

**Abstract.** Abdominal aortic aneurysm (AAA) is a life-threatening condition characterized by permanent enlargement of the aorta, often detected incidentally during imaging for unrelated conditions. Current management relies primarily on aneurysm diameter and growth rate, which may not reliably predict patient-specific rupture risk. Computation of AAA wall stress and strain has the potential to improve individualized risk assessment, but these analyses depend on image-derived geometry, which is subject to segmentation uncertainty and lacks a definitive ground truth for the wall boundary. While the effect of geometric uncertainty on wall stress has been studied, its influence on wall strain remains unclear.

In this study, we assessed the impact of geometric uncertainty on AAA wall strain computed using deformable image registration of time-resolved 3D computed tomography angiography (4D-CTA). Controlled perturbations were applied to the wall geometry along the surface normal, parameterized by the standard deviation for random variation and the mean for systematic inward or outward bias, both scaled relative to wall thickness.

Results show that uncertainties in AAA wall geometry reduce the accuracy of computed strain, with inward bias (toward the blood lumen and intraluminal thrombus) consistently causing greater deviations than outward bias (toward regions external to the aortic wall). Peak strain is more sensitive but less robust, whereas the 99th percentile strain remains more stable under perturbations. We concluded that, for sufficiently accurate strain estimation, geometric uncertainty should remain within one wall thickness (typically 1.5 mm).

## 1 Introduction

An abdominal aortic aneurysm (AAA) is a permanent aortic dilation, usually asymptomatic and often diagnosed incidentally during imaging for unrelated conditions. If untreated, AAA may expand and rupture, often fatally [1, 2]. AAA management is currently based on maximum diameter and growth rate, with intervention recommended when the aneurysm exceeds 5.5 cm in men, 5 cm in women, or grows more than 1 cm per year [2]. However, maximum diameter alone may not reliably predict rupture risk, as some AAAs rupture below the threshold while others remain stable above it [1-4].

For personalized AAA management, AAA biomechanics, particularly wall stress and tension, have been widely studied [5-8]. Stress-based rupture risk indicators rely on assumed wall strength, derived from population data due to the lack of patient-specific values [9, 10], undermining their robustness and clinical utility. Recently, non-invasive in vivo AAA wall strain measurements using sequential images from different phases of the cardiac cycle, known as 4D imaging [11], have been proposed [12-14]. More recently, deformable image registration of time-resolved 3D computed tomography angiography (4D-CTA) [11] has enabled patient-specific, in vivo, non-invasive AAA strain computation [12], facilitating the structural integrity assessment of the AAA wall [15].

Previous studies have shown that variations in AAA geometry influence computed wall stress and tension [16-18], but their effect on wall strain has not yet been investigated. This study addresses this gap by examining how uncertainties in AAA geometry affect computed wall strain from 4D-CTA using deformable image registration. The geometric variations considered here are not intended to represent differences between patients, but rather to reflect segmentation-related uncertainty for a single individual. Using 4D-CTAs and segmentations of AAAs from the publicly available dataset [11], we apply controlled perturbations to AAA wall geometry to assess the impact of geometric uncertainty on the magnitude and distribution of wall strain.

The remainder of this paper is organized as follows. Section 2 presents the patient-specific AAA imaging and geometry data, together with a summary of the strain computation approach and the geometric perturbation method. In Section 3, we report the results on the extent of changes in strain magnitude and distribution as a function of geometric perturbations, followed by conclusions and discussion in Section 4.

## 2 Materials and methods

### 2.1 Image Data and Patient-Specific Geometry of Abdominal Aortic Aneurysm (AAA)

The image and geometry data used in this study were obtained from our publicly available 4D-CTA dataset of AAAs [11]. For each patient, the dataset includes ten electrocardiogram (ECG)-gated 3D-CTA image frames, captured uniformly at 10% intervals of the R–R interval across the cardiac cycle. In this dataset, patient-specific AAA geometries were extracted from the images using automated AI-based segmentation for image labeling, and the freely available BioPARR (Biomechanics-based Prediction of Aneurysm Rupture Risk) software [6] for generating AAA surface models, which were provided as triangulated surface representations. From this dataset, we selected patients 1, 2, 3, 7, and 8, which are referred to as Patients 1 to 5 in this paper for consistency.

As an example, Fig. 1 shows the 3D-CTA image in the systolic phase (Fig. 1a) together with the external wall geometry (Fig. 1b) of Patient 1's AAA in the patient-specific coordinate system, where the L-R (left-right), P-A (posterior-anterior), and I-S (inferior-superior) axes are indicated. The data were visualized using the 3D Slicer image computing platform [19]. As shown in Fig. 1c on a selected R–A plane, the external wall contours delineate the outer boundary while remaining within the high-gradient fuzzy region of the AAA wall, where a precise wall location cannot be clearly defined. It should be noted that, to the best of our knowledge, there is no universally accepted definition or ground truth for the exact location of the AAA wall. As described in Section 2.1, the wall location used in this study was obtained from AI-based segmentation performed with a commercial software application. The triangulation vertices define a point cloud representation of the AAA wall (Fig. 1d), which serves as the input geometry for strain analysis. The point cloud shown in Fig. 1d consists of 29,661 points.

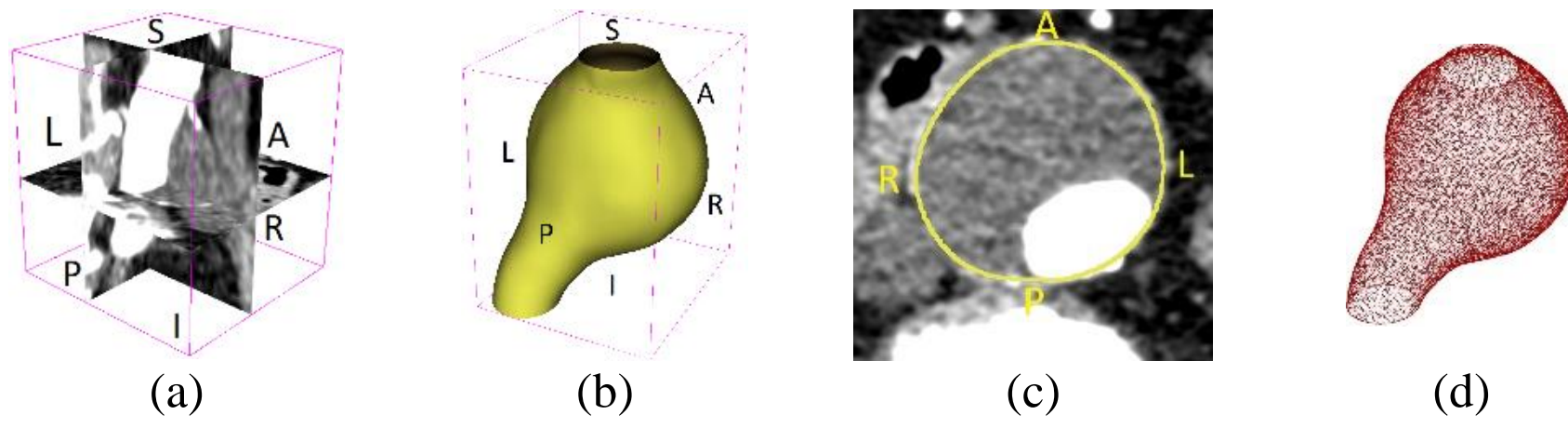

**Fig. 1.** Patient 1's abdominal aortic aneurysm (AAA): (a) 3D-CTA image in the systolic phase, (b) external wall surface model, (c) AAA external wall contours on a selected R-A plane, and (d) point cloud representation of the AAA external wall surface.

### 2.2 Image registration and strain computation

Building on our recent research on AAA kinematics [12], we used deformable image registration to align systolic (fixed) and diastolic (moving) 3D volumes of 4D-CTA and to estimate the displacement field mapping systolic to diastolic AAA geometry. We calculated wall displacement in the Cartesian patient coordinate system (R, A, S) by interpolating the registration displacement field at the point cloud representing the AAA external wall surface (Fig. 1d).

Previous studies have shown that registration provides more accurate displacement estimation along the image gradient, which, in the AAA wall region of 3D-CTA images, closely aligns with the wall normal [12, 20]. Therefore, using 3D plane fitting, we established a local biological coordinate system consisting of the local wall normal and two perpendicular tangential directions, and decomposed the wall displacement into its normal and tangential components [12].

Following [12, 15, 21, 22] and considering the predominance of wall circumferential strain [23, 24], we used the wall normal displacement from registration (ΔR) to compute the local circumferential wall strain as [25, 26]:

$$\epsilon = \frac{\Delta R}{R} \tag{1}$$

where $R$ represents the local radius of wall curvature, estimated through local surface fitting with non-deterministic outlier detection [12, 27]. We built upon the *imregdeform* deformable image registration algorithm available in the MATLAB programming language, with isotropic total variation regularization of displacement, to implement the strain calculation methods [12, 28]. For implementation details and parameter settings, see [12].

### 2.3 Analysis of the effects of geometric uncertainty

To assess the impact of geometric uncertainty on the magnitude and distribution of AAA wall strain, the AAA point cloud in Fig. 1d was taken as the "reference" AAA wall geometry, which was then perturbed. A Gaussian random noise was used to prescribe independent random deviations along the local surface normal at each point of the wall external surface [18, 29] . The perturbations were parameterized by two quantities: the standard deviation ($\sigma$), representing the perturbation magnitude, and the mean ($\mu$), representing a potential bias (mean deviation) in either the inward or outward direction. A zero-mean case ($\mu = 0$) corresponds to unbiased noise centered on the reference wall, while non-zero mean values capture systematic bias. A positive mean indicates an outward bias, whereas a negative mean corresponds to an inward bias. To ensure physiologically relevant perturbations, both $\sigma$ and $\mu$ were scaled as multiples of the typical AAA wall thickness of 1.5 mm, as reported in the literature [30].

We computed the strain in the perturbed wall using the same procedure as for the reference wall (ground truth strain), i.e., by interpolating the registration displacement field on the perturbed wall point cloud and calculating the local strain at each point as the ratio of normal displacement to the radius of curvature. The radius of curvature at each point was taken from the reference wall, and the computed strain in the perturbed wall was plotted on the reference (unperturbed) wall point cloud to ensure consistent visualization and allow comparison of strain fields between the reference and perturbed walls.

This formulation is practically important because, in the absence of ground truth for wall segmentation, variations occur in the position of the external wall across the wall thickness. By systematically varying $\sigma$ and $\mu$, we evaluated how geometric uncertainty affects the magnitude and distribution of computed wall strain.

## 3 Results

Fig. 2 presents the computed AAA wall strain in the perturbed wall of Patient 1 for different combinations of perturbation magnitude (σ) and bias (μ). The cases include unbiased perturbations ($\mu = 0$) with $\sigma = 0$ mm (reference wall geometry), $\sigma = 1.5$ mm, and $\sigma = 3.0$ mm, as well as biased perturbations (with perturbation magnitude $\sigma = 1.5$ mm) using $\mu = +3.0$ mm (outward bias) and $\mu = -3.0$ mm (inward bias). For each case, the perturbed wall geometry, the strain distribution in the perturbed wall, and the difference in strain between the perturbed and reference walls are shown. The first row in Fig. 2 represents the reference wall geometry, with its strain considered as the ground truth strain. Fig. 2 shows that strain deviation from the ground truth generally increases with both perturbation magnitude and bias. Larger unbiased perturbations result in greater differences in strain between the perturbed and reference walls. Inward and outward biases also produce distinct strain maps, with inward bias ($\mu$ = –3 mm) causing greater strain deviation than outward bias ($\mu$ = +3 mm) for the same perturbation magnitude ($\sigma$ = 1.5 mm).

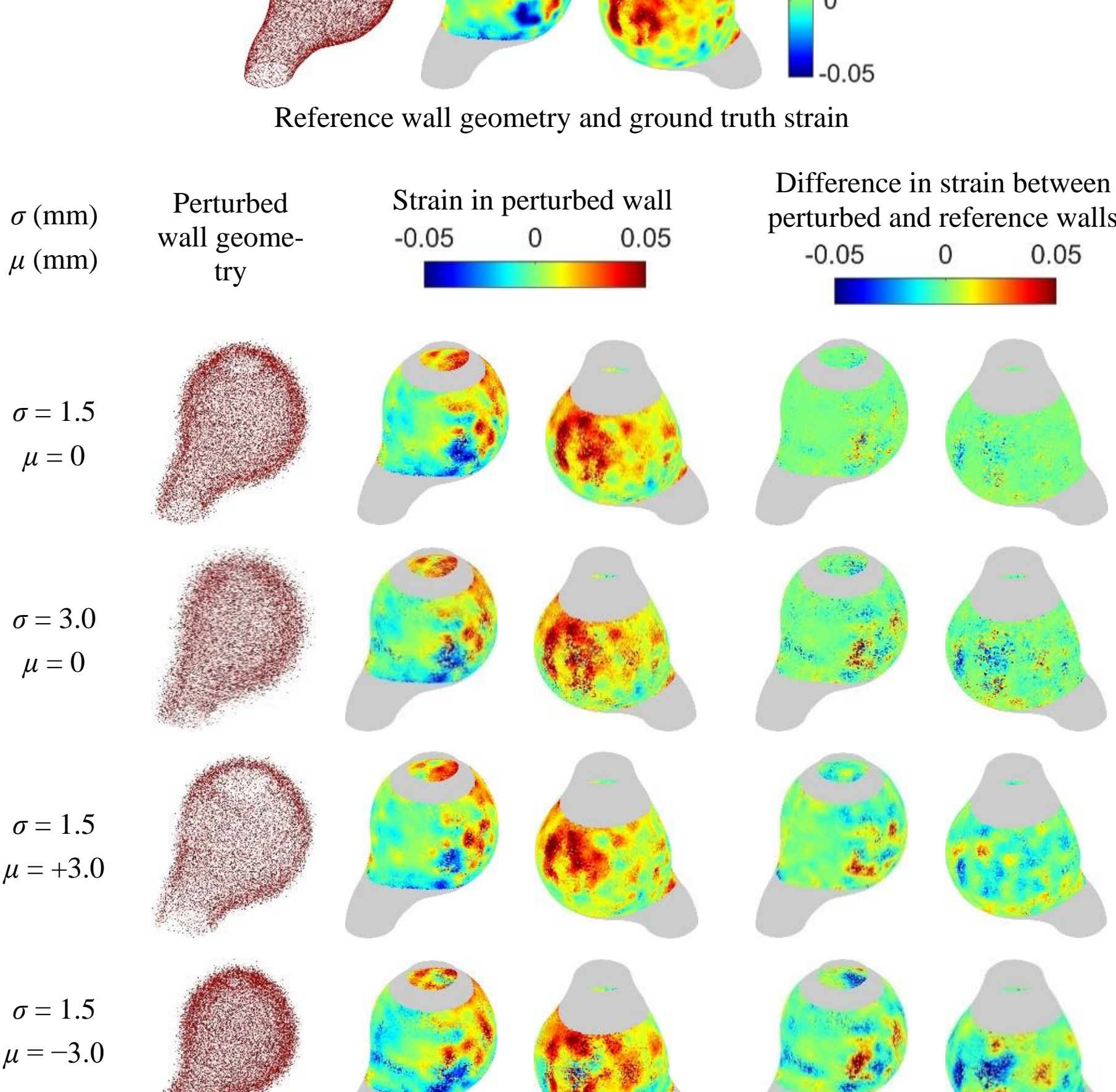

**Fig. 2.** Strain computation in the perturbed wall of Patient 1's abdominal aortic aneurysm (AAA). The table shows the perturbed wall geometry, strain in the perturbed wall, and the difference in strain between the perturbed and reference walls for various values of perturbation magnitude ($\sigma$) and bias ($\mu$). Positive mean ($\mu > 0$) indicates outward bias, whereas negative mean ($\mu < 0$) indicates inward bias. The first row represents the reference wall geometry, with its strain considered as the ground truth strain. We analyzed strain only within the AAA wall and excluded the transition zone between the AAA and the healthy aorta, shown in grey, from the analysis.

To quantify the impact of geometric perturbations on AAA wall strain distribution, Fig. 3 presents scatter plots of strain in the perturbed wall versus the ground truth strain for different combinations of perturbation magnitude ($\sigma$) and bias ($\mu$), for the cases in Fig. 2: $\sigma$ = 1.5 mm, $\mu$ = 0 mm (Fig. 3a); $\sigma$ = 3.0 mm, $\mu$ = 0 mm (Fig. 3b); $\sigma$ = 1.5 mm, $\mu$ = +3.0 mm (Fig. 3c); and $\sigma$ = 1.5 mm, $\mu$ = –3.0 mm (Fig. 3d). The coefficient of determination ($R^2$) and the Normalized Root Mean Square Error (NRMSE) for the identity line fit (y = x) are shown in each scatter plot to quantify deviations from the ground truth strain.

To strengthen and broaden the applicability of our results and subsequent discussion, we applied the strain computation in perturbed wall geometry to four additional patients, as shown in Fig. 4. This figure shows the point cloud representation of the AAA wall for Patients 2 to 5 together with the corresponding strain maps.

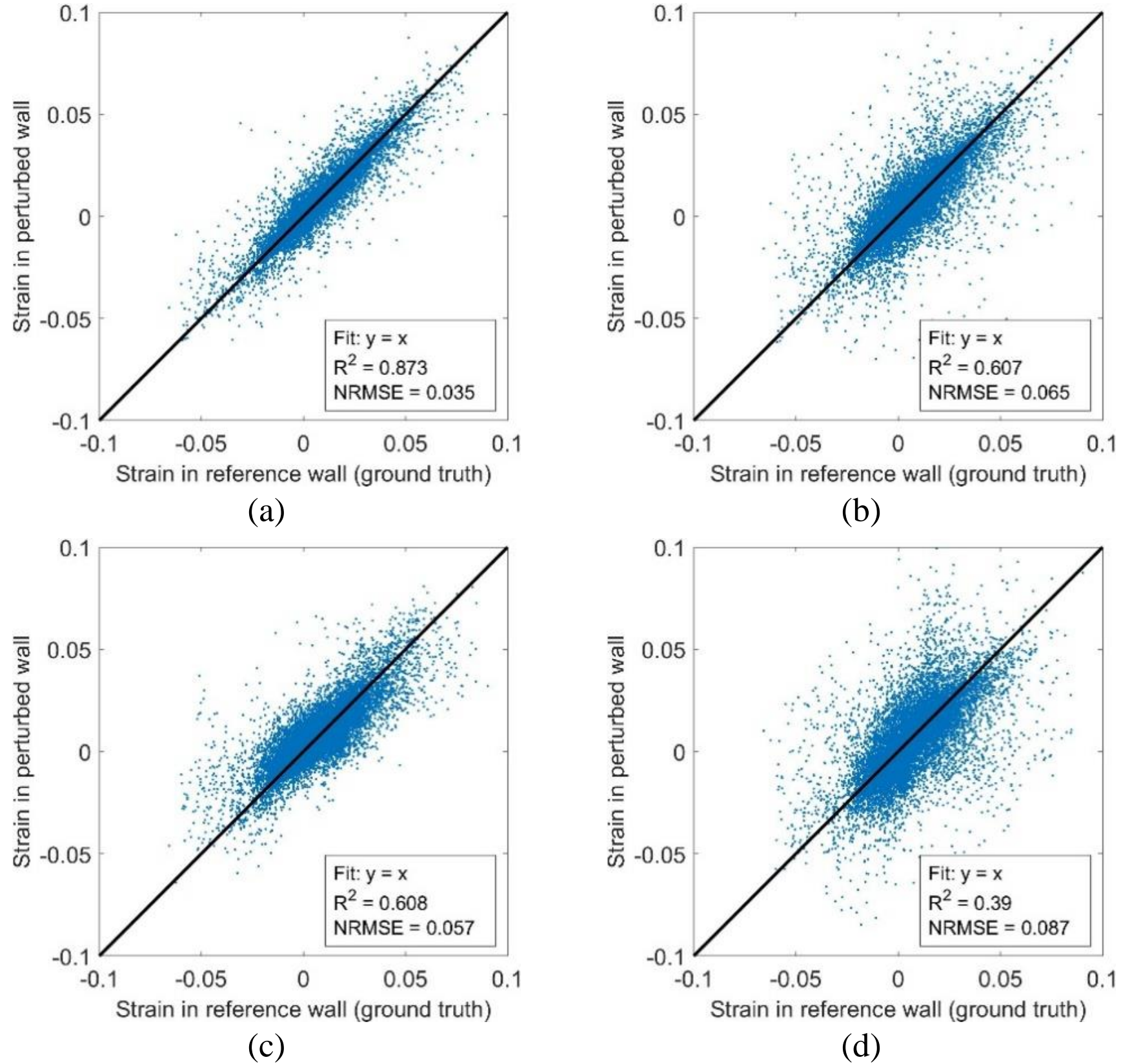

**Fig. 3.** Scatter plots of strain in the perturbed wall geometry versus the ground truth strain (strain in the reference wall geometry) for Patient 1, shown for different combinations of perturbation magnitude ($\sigma$) and bias ($\mu$): (a) $\sigma$ = 1.5 mm, $\mu$ = 0 mm; (b) $\sigma$ = 3.0 mm, $\mu$ = 0 mm; (c) $\sigma$ = 1.5 mm, $\mu$ = +3.0 mm; and (d) $\sigma$ = 1.5 mm, $\mu$ = −3.0 mm. Positive mean ($\mu > 0$) indicates outward bias, whereas negative mean ($\mu < 0$) indicates inward bias. Each plot shows the coefficient of determination ($R^2$) and the Normalized Root Mean Square Error (NRMSE) for the identity line fit (y = x).

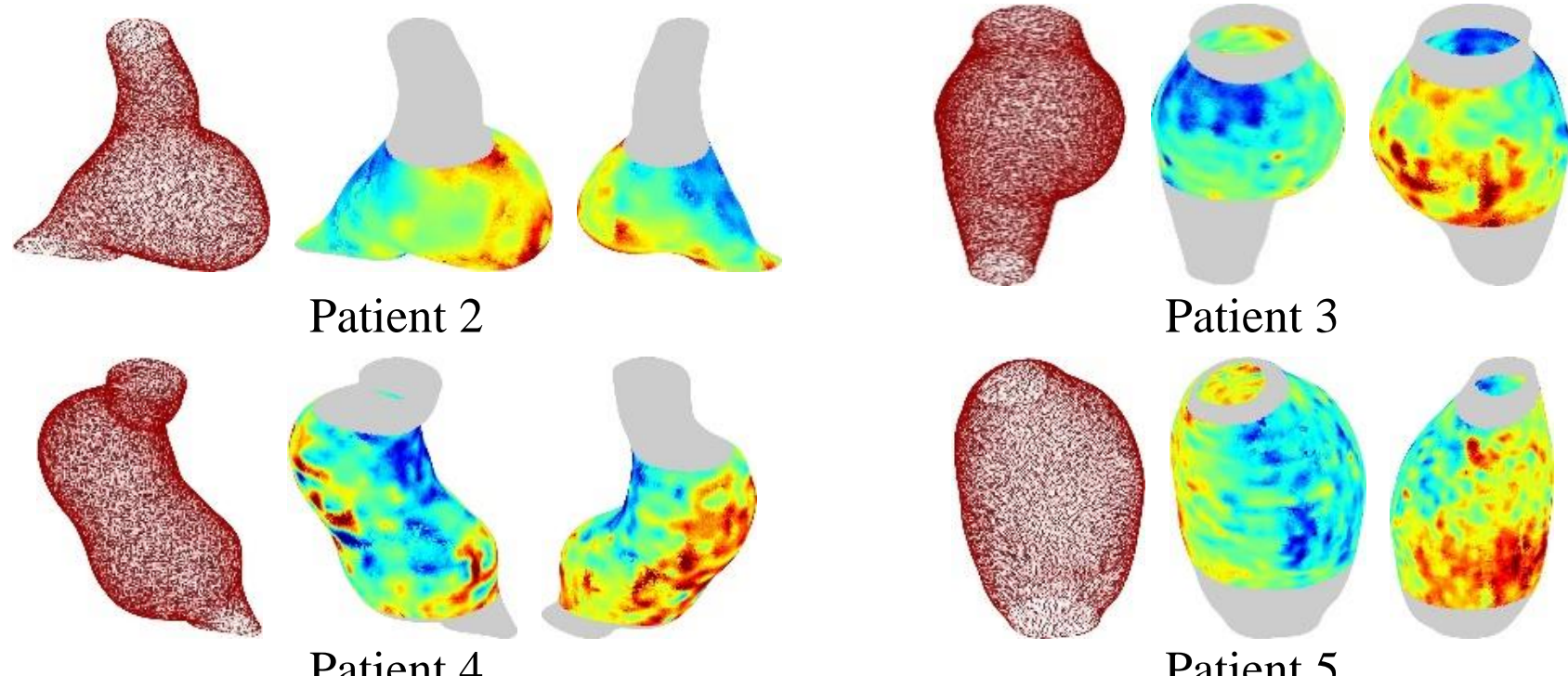

**Fig. 4.** Point cloud representation of the AAA wall for Patients 2 to 5 and the corresponding strain maps. We analyzed strain only within the AAA wall and excluded the transition zone between the AAA and the healthy aorta, shown in grey, from the analysis.

To provide a broad view of the effects of geometric perturbations on AAA wall strain distribution, Fig. 5 shows the variations of $R^2$ and NRMSE as functions of perturbation magnitude ($\sigma$) and mean ($\mu$), respectively, across all five patients analyzed in this study. For each patient, and for each prescribed $\sigma$ and $\mu$, $R^2$ and NRMSE are computed for the identity line fit (y = x) to the strain in the perturbed wall versus the ground truth strain. Three distinct perturbation scenarios are analyzed, as shown in Fig. 5.

Fig. 5a corresponds to unbiased perturbations with $\mu = 0$ and $\sigma$ ranging from 0 to 9 mm in increments of 1.5 mm. Fig. 5b corresponds to pure offset perturbations with $\sigma = 0$ and $\mu$ varying between –9 mm (inward bias) and +9 mm (outward bias) in increments of 1.5 mm. Fig. 5c corresponds to biased perturbations with $\sigma$ fixed at 1.5 mm and $\mu$ varying between –9 mm and +9 mm, also in increments of 1.5 mm. Given a typical wall thickness of 1.5 mm, the perturbation magnitude spans up to six wall thicknesses, and the bias range corresponds to inward and outward offsets of up to six wall thicknesses, both with increments equal to one wall thickness.

$R^2$ and NRMSE quantify how closely the strain in the perturbed wall aligns with the ground truth strain. $R^2$ measures the strength of the linear relationship, with values near 1 indicating strong correlation and minimal deviation, while values near 0 suggest weak correlation and significant deviations due to geometric perturbations. NRMSE quantifies the overall error between the perturbed wall strain and the ground truth strain, with lower values indicating closer agreement and higher values indicating greater deviation.

The results reported in Fig. 5 show that $R^2$ decreases and NRMSE increases as perturbation magnitude ($\sigma$) or mean ($\mu$) increases, indicating that the correlation between strain in perturbed walls and the ground truth strain weakens with larger perturbations. While the rate of deterioration of $R^2$ and NRMSE with increasing geometric perturbations varies across patients (for example, Patient 2 shows the least sensitivity compared to the others) the overall trends are consistent across all patients.

Fig. 5 shows that satisfactory agreement, defined by $R^2 > 0.8$ and NRMSE $< 0.05$, is achieved for all patients when perturbation magnitude and mean remain within one wall thickness, that is, when $\sigma < 1.5$ mm and $|\mu| < 1.5$ mm. Comparing Fig. 5a and Fig. 5b indicates that perturbation bias, rather than perturbation magnitude, has a more pronounced effect on strain deviation. Furthermore, Fig. 5c demonstrates that the combined effect of bias and magnitude produces greater strain deviation than either factor alone, with inward bias (toward the blood lumen and intraluminal thrombus) generally causing larger deviations than outward bias (toward regions external to the aortic wall).

Overall, the results in Fig. 5 indicate that uncertainties in AAA wall geometry, particularly systematic underestimation (inward bias), reduce the accuracy of computed wall strain. For accurate estimation of AAA wall strain distribution, geometric perturbation magnitude should be limited to less than one wall thickness, and geometric perturbation bias should be restricted to no more than one wall thickness inward or outward from the reference geometry, (i.e. the actual AAA geometry as depicted in the CT images.

To provide a comprehensive view of the influence of geometric perturbations on key numerical indicators of AAA wall strain proposed in the literature [12, 31], Fig. 6 presents the variations of peak wall strain and 99th percentile wall strain as functions of perturbation magnitude ($\sigma$) and perturbation mean ($\mu$), across all five patients. In each case, the peak and 99th percentile values are extracted from the strain distribution in the perturbed wall geometry defined by the prescribed $\sigma$ and $\mu$. Consistent with Fig. 5, three distinct perturbation scenarios are examined.

Fig. 6a corresponds to unbiased perturbations ($\mu = 0$) with $\sigma$ varying between 0 and 9 mm in increments of 1.5 mm. Fig. 6b corresponds to pure offset perturbations ($\sigma = 0$) with $\mu$ ranging from –9 mm (inward bias) to +9 mm (outward bias) in increments of 1.5 mm. Fig. 6c corresponds to biased perturbations with $\sigma$ fixed at 1.5 mm and $\mu$ ranging from –9 mm to +9 mm, also in increments of 1.5 mm.

Fig. 6 demonstrates that both peak and 99th percentile wall strain deviate further from the ground truth as perturbation magnitude ($\sigma$) or bias ($\mu$) increases, with peak strain showing greater sensitivity to perturbations than the 99th percentile strain.

In Fig. 6a, which corresponds to unbiased perturbations ($\mu = 0$), peak strain remains relatively stable for perturbation magnitudes up to one wall thickness ($\sigma < 1.5$ mm), whereas the 99th percentile strain shows only minor variations up to two wall thicknesses ($\sigma < 3.0$ mm).

For pure offset perturbations ($\sigma = 0$), both peak and 99th percentile strain deviate progressively as the bias ($\mu$) increases in either direction (Fig. 6b). The deviations are more pronounced for inward bias ($\mu < 0$, toward the blood lumen and intraluminal thrombus) than for outward bias ($\mu > 0$, toward regions external to the aortic wall). Similar to the results in Fig. 6a, peak strain is more sensitive to mean perturbations than the 99th percentile strain, showing larger fluctuations across patients.

Notably, Fig. 6b suggests a relatively stable point for the 99th percentile strain when $\mu$ is between +1.5 mm and +3 mm, indicating that if the reference wall geometry were offset by this amount, the 99th percentile strain would be only slightly affected by perturbation bias within one wall thickness. In contrast, the peak strain shows substantial fluctuations, even for the bias within one wall thickness ($\mu < 1.5$ mm), highlighting its greater sensitivity.

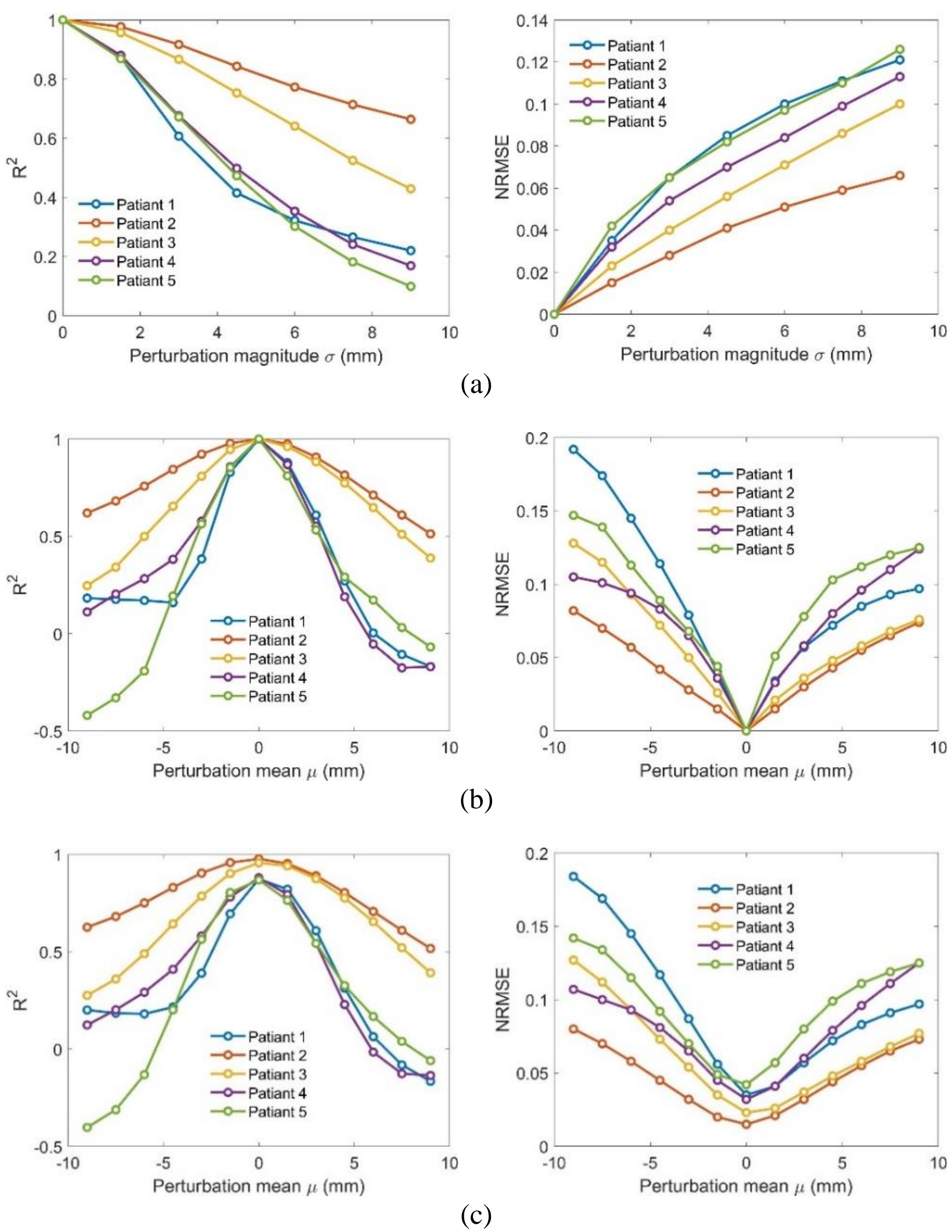

**Fig. 5.** Coefficient of determination ($R^2$) and Normalized Root Mean Square Error (NRMSE) of the identity line fit ($y = x$) to scatter plots of strain in perturbed walls versus ground truth strain for Patients 1 to 5, as functions of: (a) perturbation magnitude ($\sigma$) with mean $\mu = 0$ (unbiased perturbations), (b) perturbation mean ($\mu$) with fixed perturbation magnitude $\sigma = 0$ (pure offset inward or outward), and (c) perturbation mean ($\mu$) with fixed perturbation magnitude $\sigma = 1.5$ mm (biased perturbations). Positive mean ($\mu > 0$) indicates outward bias, whereas negative mean ($\mu < 0$) indicates inward bias.

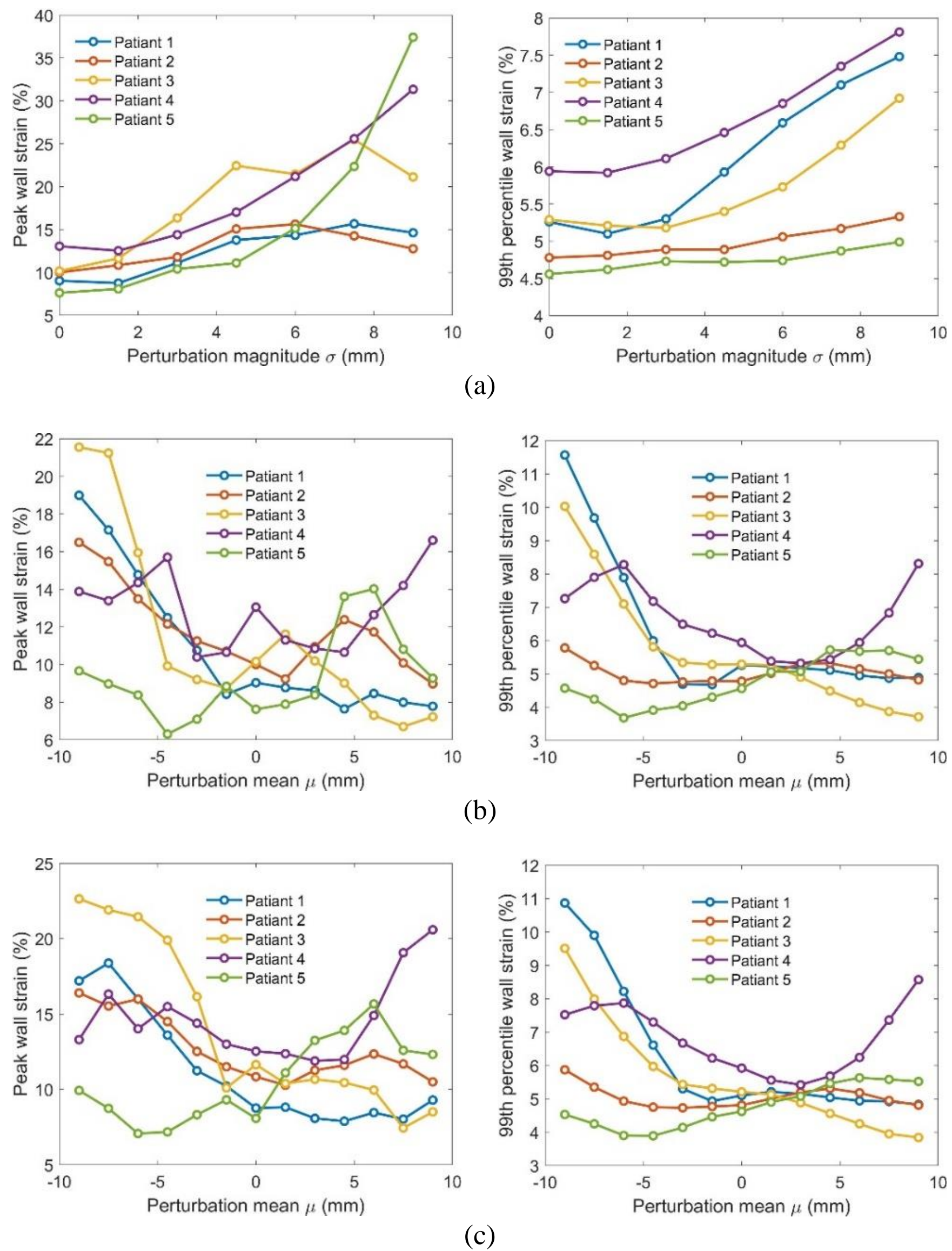

**Fig. 6.** Peak and 99th percentile wall strain in the perturbed wall for Patients 1 to 5, as functions of: (a) perturbation magnitude ($\sigma$) with mean $\mu = 0$ (unbiased perturbations), (b) perturbation mean ($\mu$) with fixed perturbation magnitude $\sigma = 0$ (pure offset inward or outward), and (c) perturbation mean ($\mu$) with fixed perturbation magnitude $\sigma = 1.5$ mm (biased perturbations). Positive mean ($\mu > 0$) indicates outward bias, whereas negative mean ($\mu < 0$) indicates inward bias.

As indicated in Fig. 6c, which shows the results for biased perturbations with $\sigma$ of 1.5 mm, the combined effect of perturbation magnitude and bias further amplifies strain deviations. Both peak and 99th percentile strain deteriorate as $\mu$ increases in either direction, with the deviations being more pronounced for inward bias ($\mu < 0$) than for outward bias ($\mu > 0$). As in Fig. 6a and Fig. 6b, peak strain exhibits greater sensitivity than the 99th percentile strain, reflecting its higher susceptibility to geometric uncertainties.

The results in Fig. 5 and Fig. 6 indicate that the peak strain, rather than 99$^{th}$ percentile strain, is a more sensitive and less robust indicator of AAA wall kinematics. The 99th percentile strain offers a more stable measure under geometric uncertainties, with inward bias consistently producing the greatest deviations. For perturbations within one wall thickness, the 99th percentile strain remains relatively stable across patients, while peak strain still shows noticeable variability.

## 4 Discussion and conclusions

We investigated how geometric uncertainty affects computed wall strain in Abdominal Aortic Aneurysms (AAAs) derived from 4D-CTA image data using deformable image registration [12]. We used image and geometry data for five patients from a publicly available 4D-CTA AAA dataset [11].

The reference wall geometry was represented as a point cloud defined by the vertices of the triangulated external surfaces of the AAA walls. This geometry was systematically perturbed using a Gaussian random noise that introduced independent deviations along the local surface normal. Perturbations were characterized by two parameters: the standard deviation ($\sigma$), representing perturbation magnitude, and the mean ($\mu$), representing systematic inward or outward bias. Both parameters were scaled relative to the AAA wall thickness of 1.5 mm, often used in the literature [30].

We compared the strain distribution in the perturbed wall with that in the reference wall (ground truth) using scatter plots (Fig. 3). Quantitative analysis using the coefficient of determination ($R^2$) and the Normalized Root Mean Square Error (NRMSE) of the identity line fit (y = x), where x represents the ground truth strain and y the strain in the perturbed geometry, revealed clear and consistent trends across patients under different perturbation scenarios.

$R^2$ decreased and NRMSE increased as perturbation magnitude ($\sigma$) or bias ($\mu$) increased, with inward bias (toward the blood lumen and intraluminal thrombus) generally causing greater deviations than outward bias (toward regions external to the aortic wall). Bias had a more pronounced effect than perturbation magnitude, and the combined effect of the magnitude and bias amplified deviations from the reference strain, particularly under inward bias. This asymmetry can be explained by the fact that image registration typically achieves more accurate displacement estimation in regions with strong image gradients or texture [12]. Inward perturbations move the wall toward the blood lumen or intraluminal thrombus, where image texture is weak, reducing registration accuracy and amplifying deviations from the ground truth strain. In contrast, outward perturbations shift the wall toward surrounding tissues with stronger texture, where registration performs more reliably, resulting in smaller errors for the same perturbation magnitude. Satisfactory agreement ($R^2 > 0.8$ and NRMSE < 0.05) between the strain in the perturbed geometry and the ground truth strain was maintained when $\sigma$ and $|\mu|$ remained within one wall thickness (1.5 mm). Therefore, it can be suggested that to ensure accurate strain estimation, uncertainties in determining the AAA wall geometry (i.e., segmentation uncertainties) should be limited to less than one wall thickness in both magnitude and bias.

We also analyzed peak and 99th percentile wall strain as key indicators of AAA wall strain. Our results showed that both measures deviated progressively with increasing AAA wall geometry perturbation magnitude ($\sigma$) and bias ($\mu$), with the peak strain being more sensitive to geometric uncertainty than the 99th percentile strain. Inward bias consistently produced the largest deviations, whereas the 99th percentile strain remained relatively stable across patients for perturbations within one wall thickness. Overall, these findings indicate that peak strain is a less robust indicator than the 99th percentile strain under typical segmentation uncertainties.

In summary, geometric uncertainty, particularly inward bias, substantially reduces the accuracy of computed AAA wall strain. While peak strain was highly sensitive to perturbations, the 99th percentile strain provided a more stable measure under segmentation uncertainties. To ensure satisfactory strain distribution and stability of the 99th percentile strain, geometric uncertainty should remain within one wall thickness.

**Acknowledgments.** This work was supported by the Australian National Health and Medical Research Council NHMRC Ideas grant No. APP2001689. This study uses time-resolved computed tomography (4D-CTA) images and segmentations of abdominal aortic aneurysms (AAAs) from the publicly available dataset [11]. The authors acknowledge the contributions of Christopher Wood and Jane Polce, radiology technicians at the Medical Imaging Department, Fiona Stanley Hospital (Murdoch, Western Australia), to patient image acquisition, and Dr. Farah Alkhatib's (The University of Western Australia) assistance in patient recruitment and image acquisition.